\documentclass[aps,prd,superscriptaddress,longbibliography,amssymb,twocolumn]{revtex4-1}

\usepackage[utf8]{inputenc}
\usepackage[pdftex]{graphicx}
\usepackage{amsmath}
\usepackage[amssymb,Gray]{SIunits}
\usepackage{textcomp}
\usepackage{physics}
\usepackage{amssymb}
\usepackage{mathtools}
\usepackage{tikz}
\usepackage{hyperref}
\usepackage{float}
\usepackage[most]{tcolorbox}

\usepackage{xcolor}
\newcommand{\mn}{\mu \nu}

\newcommand{\ri}{\textbf{r}_{\mathrm{I}}}

\newcommand{\vi}{\dot{\textbf{r}}_{\mathrm{I}}}

\newcommand{\ei}{e_{\mathrm{I}}}

\newcommand{\Anp}{\mathcal{A}}
\newcommand{\nn}{\nonumber}

\renewcommand{\vec}[1]{\mathrm{\mathbf{#1}}}

\begin{document}

%Title of paper
\title{The Hamiltonian for an atom interacting with gravitational waves}

\author{Linda M. van Manen}
\email[]{linda.van.manen@uni-jena.de}
\author{Andr\'e Gro{\ss}ardt}
\email[]{andre.grossardt@uni-jena.de}
\affiliation{Institute for Theoretical Physics, Friedrich Schiller University Jena, Fr\"obelstieg 1, 07743 Jena, Germany}

\date{\today}

% PRL: abstract max. 600 characters
\begin{abstract}
Building on the relativistic Hamiltonian of Sonnleitner and Barnett \cite{BarnettSonnleitner2018} and its post-Newtonian extensions by Schwartz and Giuilini \cite{SchwartzGiulini2019}, we investigate composite atomic systems in dynamical gravitational backgrounds. Using a local inertial frame and a perturbed Minkowski metric, we derive curvature-dependent corrections to both center-of-mass and internal Hamiltonians for atoms interacting with weak gravitational waves. The resulting Hamiltonian contains distinct curvature couplings modifying the internal potential and affecting the center-of-mass dynamics. These contributions imply that internal-energy variations do not always reduce to mass renormalization and can induce genuine forces due to changes in momentum. The initial research was motivated by anomalous friction-like forces emerging in quantum optics, and clarified that the anomalous forces are mere relativistic corrections from mass-energy equivalence. Our results suggest that, with increasingly sensitive detectors, additional forces from gravitational wave interactions may become visible in future experiments.
\end{abstract}

\maketitle
\section{Introduction}

Galilean transformations relate inertial frames differing by a constant relative velocity, implying identical accelerations and, with equal masses, identical forces must be observed in all inertial frames. However, an emitting or absorption atom is at rest in its instantaneous comoving frame, but shows a change in momentum due to recoil in the laboratory frame. This suggests a friction-like force that is absent in the comoving description \cite{sonnleitner2018, SonnleitnerTrautmann2017,BarnettSonnleitner2018}. The apparent inconsistency illustrates the inadequacy of a strictly Galilean framework for atom-photon interactions. The resolution lies in recognizing that photon emission reduces the atom’s internal energy \cite{BarnettSonnleitner2018}. Through Einstein’s relation $E = mc^2$, this corresponds to a decrease in inertial mass and hence to a momentum change via $\dot{\vec{P}}=\dot{m}\vec{v}$, instead of $\dot{\vec{P}}= m \ddot{\vec{R}}$. 

The issue is often removed by ad hoc reasoning in nonrelativistic quantum-optics treatments. However, such reasoning risks obscuring the underlying physics. After all, classical intuition and nonrelativistic approximations may not apply or can be misleading when nonclassical or relativistic properties dominate the behavior of composite quantum systems. A consistent description demands explicit calculations ensuring that composite-system Hamiltonians clearly distinguish between changing momentum connected to (nonrelativistic ) acceleration or from relativistic effects.   

Sonnleitner and Barnett \cite{sonnleitner2018} addressed this by deriving a minimally relativistic Hamiltonian incorporating mass-energy equivalence,
\begin{equation}
    H = \frac{\mathbf{P}^2}{2M}\left(1 - \frac{H_A}{Mc^2}\right) + \dots ,
\end{equation}
obtained from the replacement $M \rightarrow M + E_{\rm int}/c^2$. This shows that inertia depends on the internal Hamiltonian $H_A$. The omitted terms describe the radiation field and atom-light interaction. Importantly, this approach distinguishes between momentum changes produced by acceleration and those purely associated with internal dynamics, a separation not evident in Hamiltonian formalism in standard nonrelativistic treatments.

Schwartz and Giulini \cite{SchwartzGiulini2019,schwartz2020} extended this framework by incorporating post-Newtonian corrections, producing a consistent Hamiltonian for quantum-optical systems in static gravitational fields. The results are relevant for Earth-based experiments. Complementary work by Zych et al. \cite{Zych2019} demonstrated that anomalous couplings in Coulomb systems on post-Newtonian backgrounds arise from using distant-observer coordinates. When expressed in the local rest frame, these reduce to gravitational-redshift effects. The results fit into a broader effort to understand how gravity couples to quantum systems within Einstein’s equivalence principle \cite{Zych2018, Zych2019}. Taken all together, these developments underscore the need for careful treatment of internal dynamics in relativistic contexts. 

In this work, we extend these approaches to atoms interacting with weak gravitational waves. Gravitational waves provide a genuinely dynamical environment for quantum systems, and modern interferometers, such as LIGO \cite{LIGO2009, Abbott2016}, Virgo \cite{Virgo2021}, KAGRA \cite{KAGRA2025, Akutsu2019}, and LISA \cite{LISA2025}, rely on quantum-optics techniques. They operate at quantum-limited sensitivity, motivating proposals to probe quantum aspects of gravity (e.g., signatures of Planck-scale spacetime fluctuations \cite{Amelino1999, Hogan2012}). It provides observations of relativistic, dynamical environment acting on quantum scales, and with the increasing sensitivity of quantum optics experiments in a laboratory, a theoretical framework that accurately includes the relativistic effects in such experiments is required.

We adopt the electromagnetic structure of the Sonnleitner-Barnett model in a perturbed Minkowski background and work in a local inertial frame comoving with one constituent particle. Although the affine connection vanishes along this worldline, its expansion generates curvature-dependent corrections coupling both to the particles and their internal electromagnetic fields. This is contrary to the static-field treatment of Schwartz and Giulini, where metric derivatives vanish.

\section{Classical Hamiltonian for Atom-Gravity Interaction}

We consider a Hydrogen-like atom composed of two point particles with masses $m_{\mathrm{I}}$, positions $\mathbf r_{\mathrm{I}}$, and charges $e_{\mathrm{I}}$. The particles interact via internal Coulomb and vector potentials $(\phi, \mathcal{A})$, and are coupled to an external transverse electromagnetic field $\mathbf A^\perp$. The two particles experience weak relativistic effects, i.e., $v\ll c$, and working in the Coulomb gauge, $\nabla \cdot \mathbf A = 0$, the Lagrangian to order $\order{\vec{v}^2/c^2}$ reads \cite{sonnleitner2018}
\begin{align}
L_{EM} =& \sum_{I<J}^2 \Bigg[
\frac{1}{2} m_{\mathrm{I}} v_{\mathrm{I}}^2 + \frac{m_{\mathrm{I}}}{8c^2} v_{\mathrm{I}}^4
- \frac{e_{\mathrm{I}} e_{\mathrm{J}}}{4\pi\epsilon_0 r}\left(1 - \frac{\mathbf v_{\mathrm{I}} \cdot \mathbf v_{\mathrm{J}}}{2c^2}\right) \nn \\
&+ \frac{e_{\mathrm{I}} e_{\mathrm{J}}}{8\pi\epsilon_0 c^2} \frac{(\mathbf v_{\mathrm{I}} \cdot \mathbf r)(\mathbf v_{\mathrm{J}} \cdot \mathbf r)}{r^3}
\Bigg]
\nn \\
&+ L_{\mathrm{field}} + \int d^3 r\, \mathbf j \cdot \mathbf A^\perp ,
\end{align}
where $\vec{r} = \vec{r}_{\mathrm{I}} - \vec{r}_{\mathrm{J}}$ and $r = |\vec{r}|$ is its magnitude. The subscripts $\mathrm{I}, \mathrm{J}$ indicate the particle number. The rest energy, $m_{\mathrm{I}} c^2$, is omitted since the focus is on the dynamics. The Lagrangian consists of the particle's kinetic energy, and the Darwin Lagrangian describing the interaction between two charged particles. The last line contains the free-field Lagrangian for $\mathbf A^\perp$, as well as the interaction between the source and the external field. 

The cross terms between internal and external fields originate from the back-action of external fields on the charge-generated internal potentials. Following Sonnleitner and Barnett \cite{sonnleitner2018}, these contributions, combined with divergent self-interaction terms arising in $J_{\mu} A^{\mu}$, are neglected throughout this work, as they do not contribute to the physical processes under investigation. 

\subsection{Gravitational corrections}

We introduce weak gravitational perturbations acting on the atomic kinetic sector, assuming the atom is in local free fall (e.g.\ in \cite{ESASpaceMicroscopeUFF, ZARMDropTowerBremen}). The spacetime metric is
\begin{equation}
g_{\mu\nu} = \eta_{\mu\nu} + h_{\mu\nu}, \qquad |h_{\mu\nu}|\ll 1 ,
\label{metric}
\end{equation}
and we employ Fermi-Normal coordinates to construct a local inertial frame. To quadratic order in the spatial separation $r^i = (r_{\mathrm{I}} - r_{\mathrm{J}})^i$, the metric becomes
\begin{equation}
g_{\mu\nu} =
\begin{pmatrix}
-1 - R_{0i0j} r^i r^j & -\frac{2}{3} R_{0jik} r^j r^k \\
-\frac{2}{3} R_{0jik} r^j r^k & \delta_{ij} - \frac{1}{3} R_{ikjl} r^k r^l
\end{pmatrix},
\end{equation}\\
expressed in terms of local Riemann components. The particle Lagrangian becomes
\begin{align}
    L_{kin}
    =& \sum_{\mathrm{I}}\Big[\frac{ m_{\mathrm{I}} \vec{v}_{\mathrm{I}}^2}{2} + \frac{ m_{\mathrm{I}} \vec{v}_{\mathrm{I}}^4}{8 c^2} - \frac{m_{\mathrm{I}}}{4} R_{0i0j} r^i r^j \vec{v}_{\mathrm{I}}^2 \nn \\
   &- \frac{2}{3}  m_{\mathrm{I}} c \;R_{0jik}\; r^j r^k \; v_{\mathrm{I}}^i - \frac{ m_{\mathrm{I}}}{6} R_{ikjl} \; r^k r^l\, v_{\mathrm{I}}^i v_{\mathrm{I}}^j \Big].
   \label{Kinetic-Lagrangian}
\end{align}
Note that the omitted rest energy $- m_{\mathrm{I}} c^2$ gains a correction $ - \frac{ m_{\mathrm{I}} c^2}{2} R_{0i0j}\; r^i r^j$, which is similarly omitted.
The dominant contribution is the gravito-electric potential $h_{00}= - R_{0i0j} r^i r^j$, which describes a mass renormalization: $\frac{1}{2} m(1 + \tfrac12 h_{00}) \vec{v}^2 = \frac{1}{2} m(1- \tfrac12 R_{0i0j} r^i r^j) \vec{v}^2$. The last terms in (\ref{Kinetic-Lagrangian}) encode corrections to the velocity.

\subsection{Classical Hamiltonian}

The canonical momentum obtained from the full Lagrangian (electromagnetic + gravitational) is
\begin{align}
\vec{p}_{\mathrm{I}} = & (m_{\mathrm{I}} + \delta m_{\mathrm{I}}) \vec v_{\mathrm{I}} + \sum_i m_{\mathrm{I}} v_{\mathrm{I}}^{\,j} h_{ij}
+ \frac{m_{\mathrm{I}}}{2c^2}\, \vec{v}_{\mathrm{I}}^3
 \nn \\
&+\sum_i m_{\mathrm{I}} c \, h_{0i} + e_{\mathrm{I}} \vec{A}^{\perp} 
+\frac{e_{\mathrm{I}} e_{\mathrm{J}}}{8 \pi \epsilon_0 r c^2}\left(\vec{v}_{\mathrm{J}} + \frac{\vec{r}(\vec{v}_{\mathrm{J}} \cdot \vec{r})}{r^2}\right),
\end{align}
where 
$\delta m_{\mathrm{I}} = \tfrac12 m_{\mathrm{I}} h_{00}$ modifies the inertial mass, and the $h_{ij}$ coupling modifies the momentum. The term
\begin{align}
-\tfrac{2 m_{\mathrm{I}} c}{3} \sum_i R_{0jik}\, r^j r^k 
= \sum_i m_{\mathrm{I}} c \, h_{0i},
\end{align}
is independent of the particle velocity and analogous in structure to the electromagnetic coupling $e_{\mathrm{I}} \vec{A}^{\perp}$. Introducing
\begin{equation}
\bar{\mathbf p}_{\mathrm{I}} = \mathbf p_{\mathrm{I}} - e_{\mathrm{I}} \mathbf A^\perp - m_{\mathrm{I}} c\, \mathbf h_0,
\end{equation}
with $\vec{h}_0$ the notation for the vector $h_{0i}$, a Legendre transformation yields the Hamiltonian
\begin{align}
    H =& \sum_{\mathrm{I}} \, \frac{\Bar{\vec{p}}_{\mathrm{I}}^2}{2 \tilde{m}_{\mathrm{I}}} - \frac{\Bar{\vec{p}}_{\mathrm{I}}^4}{8 \tilde{m}_{\mathrm{I}}^{\,3} c^2} + \frac{e_{\mathrm{I}} e_{\mathrm{J}}}{4 \pi \epsilon_0 r} \nn \\[4pt]
    &- \frac{ h_{ij}\, \Bar{p}_{\mathrm{I}}^{\,i} \, \Bar{p}_{\mathrm{I}}^{\,j} }{2 m_{\mathrm{I}}^{\,2}} + \frac{ h_{ij}\,\Bar{p}_{\mathrm{I}}^{\,i} \, \Bar{p}_{\mathrm{I}}^{\,j}\, \Bar{\vec{p}}^2}{2  m_{\mathrm{I}}^3 c^2}  \nn \\[4pt]
    &+ \frac{e_{\mathrm{I}} e_{\mathrm{J}}}{8 \pi \epsilon_0 r c^2 \tilde{m}_{\mathrm{I}} \tilde{m}_{\mathrm{J}}}  
      \left[ \Bar{\vec{p}}_{\mathrm{I}} \cdot \Bar{\vec{p}}_{\mathrm{J}} 
      + \frac{(\Bar{\vec{p}}_{\mathrm{I}} \cdot \vec{r})(\Bar{\vec{p}}_{\mathrm{J}} \cdot \vec{r})}{r^2}\right]\nn \\[6pt]
& - \frac{e_{\mathrm{I}} e_{\mathrm{J}} h_{ij} }{4 \pi \epsilon_0 r c^2 m_{\mathrm{I}} m_{\mathrm{J}}}  
      \left[\Bar{p}_{\mathrm{I}}^{\,i} \, \Bar{p}_{\mathrm{I}}^{\,j} 
      + \frac{(\Bar{\vec{p}}_{\mathrm{I}} \cdot \vec{r})^{i}\, (\Bar{\vec{p}}_{\mathrm{J}} \cdot \vec{r})^{\,j} }{r^2}\right],
      \label{Hamiltonian without ext}
\end{align}
with $\tilde{m} = m + \delta m$.
The Hamiltonian thus exhibits gravity-induced mass shifts and metric-dependent modifications of the kinetic energy.

\subsubsection{Approximation for slow charges}

For slowly moving charges and wavelengths large compared to atomic dimensions, the mixed components $R_{0jik}$ and spatial components $R_{ikjl}$ are suppressed by $v/c$ relative to $R_{0i0j}$. Alternatively, the gauge invariance of the Riemann tensor allows us to express the corrections in the transverse-traceless (TT) gauge \cite{maggiore, carroll2024, Misner1973}. In the TT-gauge $h_{0\mu}^{TT}=0$ and the physical gravitational degrees of freedom reside exclusively in $h_{ij}^{\mathrm{TT}}$. Under the quadrupole approximation, $h^{TT}_{ij}$ becomes solely time dependent, hence
\begin{equation}
h_{0i} \approx \partial_i \dot h^{\mathrm{TT}}_{mn} r^m r^n \approx 0,
\qquad
h_{ij} \approx \partial_i \partial_j h^{\mathrm{TT}}_{mn} r^m r^n \approx 0.
\end{equation}

The suppression of $h_{0i}$ and $h_{ij}$, however, does not extend to propagating electromagnetic radiation, which travels at the speed of light. The internal field interactions remain insensitive to these terms, since these interactions are mediated by virtual photons. For terms involving external photons, however, the coupling to $h_{0i} $ and $h_{ij}$ remains relevant. For completeness we shall derive the multipolar Hamiltonian, following Ref.~\cite{sonnleitner2018}, with $h_{0i}$ included, although this term will be omitted afterwards.

\section{Multipolar (PZW) Hamiltonian}

Following Sonnleitner and Barnett~\cite{sonnleitner2018}, the minimal–coupling Hamiltonian is transformed into its multipolar form via the Power–Zienau–Woolley (PZW) transformation~\cite{Babiker2024},
\begin{align}
&\vec p_{\mathrm{I}}  \rightarrow U \vec p_{\mathrm{I}} U^\dagger = \vec p_{\mathrm{I}} + \hbar \nabla_{\vec r}\Lambda,\\
&\Pi^\perp(\vec x)  \rightarrow \Pi^\perp(\vec x) + \mathcal P^\perp(\vec x),
\end{align}
with $\Pi$ the canonical momentum associated with the external field $\vec{A}^{\perp}$, and
\begin{equation}
U = e^{-i\Lambda} = 
\exp \left[-\frac{i}{\hbar}\int d^3x\, \mathcal P(\vec x,t)\cdot \vec A^\perp(\vec x,t)\right].
\end{equation}
Here $\mathcal P$ is the usual polarization field centered at $\vec R$. A multipole expansion yields
\begin{equation}
\hbar\nabla_{\vec r}\Lambda \approx 
\sum_{\mathrm{I}} e_{\mathrm{I}} \Big[\vec A^\perp(\vec R) 
+ \tfrac{1}{2}\vec r_{\mathrm{I}} \times(\nabla\times \vec A^\perp(\vec R))\Big],
\end{equation}
so that in the dipole approximation
\begin{equation}
\bar{\vec p}_{\mathrm{I}} \rightarrow 
\vec p_{\mathrm{I}} + \tfrac{1}{2}\vec d\times \vec B(\vec R),
\qquad 
\vec d=\sum_{\mathrm{I}} e_{\mathrm{I}} \vec r_{\mathrm{I}}.
\end{equation}

Cross terms such as 
\begin{align}
{\vec p_{\mathrm{I}} \cdot(\vec d\times\vec B(\vec R))}/{(m_{\mathrm{I}} m_{\mathrm{J}} c^2)}
\end{align}
are beyond $\mathcal O(1/c^2)$ and thus discarded, as are higher–order kinetic terms. Only the leading kinetic energy acquires a correction. The resulting multipolar Hamiltonian is

\begin{align}
H_{\rm Multi} = &
\sum_{\rm I}\left[
\frac{(\vec p_{\rm I} + \tfrac12\vec d\times\vec B(\vec R))^2}{2m_{\rm I}}
 - \frac{\vec p_{\rm I}^4}{8 m_{\rm I}^3c^2}
\right] \nonumber\\
&+ \frac{1}{2\epsilon_0}\int d^3x\, 
\big[(\Pi_\perp^2+\mathcal P_\perp^2)+\epsilon_0^2 c^2\vec B_\perp^2\big] \nonumber \\
&+ H_{\rm Coulomb} + H_{\rm Darwin} + \int d^3x\,\mathcal P\cdot \vec E^\perp,
\end{align}
with polarization density
\begin{align}
\mathcal P &= \sum_{\rm I} e_{\rm I} \int_0^1 d\lambda\, \bar{\vec r}_{\rm I}\,
\delta(\vec x-\vec R-\lambda\bar{\vec r}_{\rm I}).
\end{align}
The last integral reproduce the standard couplings
\begin{align}
\int \mathcal P\cdot \vec E^\perp 
\rightarrow -\vec d\cdot\vec E^\perp(\vec R)
+\text{higher multipoles}.
\end{align}

\subsection{Gravitational momentum transformation}

For the gravitational term $\sim mc\,h_{0i}$, an analogous unitary transformation
\begin{align}
\vec p_{\rm I} \rightarrow U_G \vec p_{\rm I} U_G^\dagger
= \vec p_{\rm I} + \hbar\nabla_{\vec r}\Gamma,
\end{align}
with
\begin{align}
U_G=\exp\left[-\frac{i}{\hbar}\int d^3x\, 
\mathcal P_G\cdot \vec h_0\right],
\end{align}
and
\begin{align}
\mathcal P_G = \sum_{\rm I} m_{\rm I} c\,\bar{\vec r}_{\rm I} 
\int_0^1 d\gamma\, 
\delta(\vec x-\vec R-\gamma\bar{\vec r}_{\rm I}),
\end{align}
produces the shift
\begin{align}
\bar{\vec p}_{\rm I} \rightarrow \vec p_{\rm I} + 
\tfrac12\,\vec d\times\vec B_G(\vec R),
\end{align}
where $\vec B_G$ is the gravito-magnetic field, given by $\vec B_G \propto \nabla \times \vec h_0$ \cite{carroll2024}. 

In defining the coordinate frame, the origin was chosen at the constituent particle location $\vec{r}_{\mathrm{I}}.$ One can choose to similarity define the polarization field centered at this point. Then
\begin{align}
\mathcal P_G = \sum_{\rm J} m_{\rm J} c\,\vec{r}
\int_0^1 d\gamma\, 
\delta(\vec x-\vec{r}_{\rm I}-\gamma \vec{r}),
\end{align}
and the multipole expansion yields $\vec d \times \vec B_G (\vec{r}_{\mathrm{J}}$), evaluated at $\vec{r}_{\mathrm{J}}$. This is inconvenient, since we aim to express the final results in center-of-mass and relative coordinates. It is also unnecessary, since $h_{0i}$ effects are minimal for slow particles. Hence, this modification may be safely neglected or approximated by evaluating $\vec B_G$ at $\vec R$. Gravitational corrections to the field momentum $\Pi^\perp$ are derived in the next section.

\section{Electromagnetic action in curved spacetime}

The action for electromagnetic fields coupled to a gravitational background, is given by \cite{Jackson1998},  
\begin{align}
        S_{EM} = \sqrt{-g} \int d^4 x \left[-\frac{1}{4 \mu_0} F_{\mu\nu}F^{\mu\nu} + A^{\mathrm{tot}}_{\mu}J^{\mu}\right],
\end{align}
where $A_{\mu}^{\mathrm{tot}} = (-\phi^{\mathrm{tot}}/c,\vec{A}^{\mathrm{tot}})$ is the four-potential of the combined (internal + external) electromagnetic field. Hence, the action naturally incorporates both the free-field dynamics and the interaction with charged particles. The four-current vector field is related to the current \emph{density} $j^{\mu}$ via 
\begin{align}
J^{\mu} = \frac{j^{\mu}}{\sqrt{-g}},
\end{align}
with $j^{\mu}$ defined for point-like particles as,
\begin{align}
    \vec{j}(\vec{r},t) &= \sum_{\mathrm{I}} \ei \vi \;\delta^{(3)}(\vec{r}-\ri(t)),\\
    \rho(\vec{r},t) &= \sum_{\mathrm{I}} \ei \;\delta^{(3)}(\vec{r}-\ri(t)).
\end{align}
The determinant of the metric expands to first order in $h_{\mu \nu}$ as
\begin{align}
\sqrt{-g} =&\, 1-\frac{1}{2}h_{00}- \frac{1}{2} \mathrm{Tr}(h_{ij})+ \mathcal{O}(h^2) \nn \\[4pt]
=&\, 1+\frac{1}{4c^2}\, \ddot{h}^{TT}_{ij} r^i r^j - \frac{1}{12} \nabla_r^2 \,h_{ij}^{TT} r^i r^j.
\end{align}
 Under the quadrupole approximation, the spatial derivatives are approximately zero, leaving  
\begin{align}
\sqrt{-g}\approx 1-\frac{1}{2}h_{00} = 1+\frac{1}{4c^2}\, \ddot{h}^{TT}_{ij}(t) \,r^i r^j.
\end{align}
Expanding the kinetic part of the electromagnetic action and retaining the first-order gravitational corrections yields
\begin{align}
    S_{EM}^{\mathrm{kin}} =&\,  \frac{\epsilon_0}{2} \int d^4 x \, \Big[(\partial_t \vec{A}^{\mathrm{tot}} + \nabla \phi)^2 \,(1+\tfrac{1}{2} h^{00}) \nn \\[4pt]
    &- h^{ij} (\partial_t A_i^{\mathrm{tot}} + \partial_i \phi)(\partial_t A_j^{\mathrm{tot}} + \partial_j \phi)\nn \\[4pt]
    &- c^2  (\nabla \times \vec{A}^{\mathrm{tot}})^2 \, (1-\tfrac{1}{2} h^{00})\nn \\[4pt]
    &+ 2 c^2 \delta^{kl} h^{ij}(\partial_i A_k^{\mathrm{tot}} - \partial_k A_i^{\mathrm{tot}})(\partial_j A_l^{\mathrm{tot}} - \partial_l A_j^{\mathrm{tot}})\nn \\[4pt]
    &- 2 c \, \delta^{ij} h^{0k} (\partial_t A_i^{\mathrm{tot}} + \partial_i \phi)(\partial_j A_k^{\mathrm{tot}} - \partial_k A_j^{\mathrm{tot}})\Big].
\end{align}
With the following relations  
\begin{align}
&\phi \nabla^2 \phi = - \frac{\rho}{\epsilon_0}\phi,\quad \nabla \times (\nabla \times \vec{A}^{\perp}) = -\nabla^2 \vec{A}^{\perp},\nn \\[4pt] 
&\vec{\mathcal{A}}^{\perp} \cdot \, \partial_t^2 \vec{\mathcal{A}}^{\perp} + (\partial_t \vec{\mathcal{A}}^{\perp})^2 = \tfrac{1}{2} \partial_t^2 (\vec{\mathcal{A}}^{\perp})^2,
\end{align}
and integrating by parts, we shall separate the action into the internal and external part. Since external scalar potentials vanish in the absence of external charges, only the internal scalar field $\phi^{\mathrm{tot}} \equiv \phi$ remains.

\subsection{The external EM field}
The action for the external electromagnetic radiation field is given by
\begin{align}
S_{\mathrm{EM}}^{\mathrm{Ext}}
 =& \, \frac{\epsilon_0}{2} \int d^4 x \;  (\partial_t \vec{A}^{\perp})^2  
    - c^2 (\nabla \times \vec{A}^{\perp})^2\nn \\[4pt]
&+ \frac{\epsilon_{0}}{4}\int d^{4}x\;
      h^{00}\Big[(\partial_{t} \vec{A}^{\perp})^{2}
                 + c^{2}\big(\nabla\times \mathbf A^{\perp}\big)^{2}\Big]\nn  \\[4pt]
& - \epsilon_{0} c \int d^{4}x\;
      h_{0i}\,\varepsilon^{ijk}\,
      (\partial_{t} A_{j})\,
      \big(\nabla\times \mathbf A^{\perp}\big)_{k}\nn  \\[4pt]
& + \frac{\epsilon_{0}}{2}\int d^{4}x\;
      h^{ij}\Big[
         -(\partial_{t} A_{i})(\partial_{t} A_{j})\nn \\[4pt]
         & \qquad - c^{2}\big(\nabla\times \mathbf A^{\perp}\big)_{i}
                 \big(\nabla\times \mathbf A^{\perp}\big)_{j}  \Big]\nn \\[4pt]
& + \frac{\epsilon_0}{4} \int d^4 x \; h^{ij} \delta_{ij} \left[ (\partial_t \vec{A}^{\perp})^2 + c^2 ( \nabla \times \vec A^{\perp})^2 \right],
\end{align}
where the last term proportional to the delta function $\delta_{ij}$ can be neglected, since $h_{ij} \delta^{ij} \sim \delta^{ij} R_{nimj} x^n x^m = R_{mn} x^m x^n$, and the Ricci tensor is zero in a vacuum background.

The canonical momentum conjugate to $A_i$ is
\begin{align}
\Pi_i =& \, \epsilon_0 \, \partial_t A_i
 + \frac{\epsilon_0}{2} h_{00}\,\partial_t A_i
 -  \frac{\epsilon_0}{2} h_{ij}\,\partial_t A^j \nn \\[4pt]
 & - \epsilon_0 c \,\varepsilon_{ijk}\,h^{0j}(\nabla\times \mathbf A^{\perp})^k ,
\end{align}
where the $h_{0i}$-dependent term, is once again a velocity-independent term. Hence, we require a field-dependent momentum shift,
\begin{align}
& \bar\Pi_i(\mathbf x)\equiv \Pi_i(\mathbf x)+F_i[\vec A](\mathbf x),\\[4pt] 
& F_i[\vec A](\mathbf x)= \epsilon_0 c\, \varepsilon_{i j k}\, h^{0j} (\nabla\times \mathbf A^{\perp})^k.
\end{align}
However, simply redefining $\Pi\mapsto\bar\Pi=\Pi+F[\vec A]$ breaks the canonical algebra. In this case, the commutator becomes
\begin{align}
[\bar\Pi_i(\mathbf x),\bar\Pi_j(\mathbf y)]
=& i\hbar \epsilon_0 c \, [h_{0j}(\vec x) \partial_{x_i} - h_{0i}(\vec x)  \partial_{x_j}]\, \delta^{(3)} (x-y),
\end{align}
which is nonzero for generic test functions. To preserve the algebra, we introduce 
\begin{align}
\Lambda[\vec A]
  \equiv \frac{\epsilon_0 c}{2}
    \int d^3x \;
    h_{0j}(\mathbf x)
    \bigl[
      \mathbf A^{\perp}(\mathbf x)\times (\nabla\times \mathbf A^{\perp}(\mathbf x))
    \bigr]^j,
\end{align}
with
\begin{align}
\frac{\delta\Lambda}{\delta A_i(\mathbf x)}
  = \epsilon_0 c \,\varepsilon^{i j k}\,
    h_{0j}(\mathbf x)\,
    (\nabla\times \mathbf A^{\perp}(\mathbf x))_k.
\end{align}
The associated unitary generator is
\begin{align}
U = \exp \left[-\frac{i}{\hbar}\Lambda[\vec A]\right],
\end{align}
which transforms the momentum into the desired form, $\Pi_i(\vec x) + F_i[\vec A](\vec x)$, i.e.,
\begin{align}
U\,\Pi_i(\mathbf x)\,U^\dagger
= \Pi_i(\mathbf x) + \frac{\delta\Lambda}{\delta A^i(\mathbf x)} 
= \bar\Pi_i(\mathbf x),
\end{align}
while preserving the algebra: \([\bar\Pi_i, \bar\Pi_j]=0 \), since the algebra is unitarily equivalent to the original one. Similarly for $\vec A$, we have $U\, A_i(\mathbf x)\,U^\dagger = \bar{A}_i (\mathbf x) = A_i (\mathbf x)$ to first order, and
\begin{align}
[\Bar{A}_i(\mathbf x),\bar\Pi_j(\mathbf y)] \; &= \; U\,[A_i(\mathbf x),\Pi_j(\mathbf y)]\,U^\dagger \nn \\
 &=\; i\hbar\,\delta_{ij}\delta^{(3)}(\mathbf x-\mathbf y),
\end{align}
and $[\Bar{A}_i, \Bar{A}_j]=0.$

Combining the particle and field generators,
\begin{subequations}
\begin{align}
\Lambda_{\mathrm{part}} =& 
- \sum_{\mathrm{I}} \ei \int_0^1 d\lambda \; \bar{\vec r}_{\mathrm{I}} \cdot \vec{A}^{\perp} (\vec R + \lambda \bar{\vec r}_{\mathrm{I}}) \nn \\
&+ m_{\mathrm{I}} c \int_0^1 d \lambda \; \bar{\vec r}_{\mathrm{I}} \cdot \vec h_0 (\vec R + \lambda \bar{\vec r}_{\mathrm{I}}),\\
\Lambda_{\mathrm{field}} =& \,
\frac{\epsilon_0 c}{2} \int d^3 x\; h_{0j}(\vec x) [\vec A^{\perp}(\vec x) \times (\nabla \times \vec A^{\perp}(\vec x)) ]^j,
\end{align}
\end{subequations}
and applying $U=\exp[-(i/\hbar)(\Lambda_{\mathrm{part}}+\Lambda_{\mathrm{field}})]$ to the Hamiltonian gives the multipolar form
\begin{widetext}
\begin{align}
H_{\mathrm{Multi}} &= \sum_{\mathrm{I}} \left[
     \frac{(\vec{p}_{\mathrm{I}} 
    + \frac{1}{2}\, \vec{d} \times \vec{B}_{\mathrm{tot}}(\vec{R}))^{\,2}}{2 m_{\mathrm{I}}}
      - \frac{\mathbf p_{\mathrm{I}}^{\,4}}{8 m_{\mathrm{I}}^3 c^2}
   \right]
   + H_{\mathrm{Coulomb}}
   + H_{\mathrm{Darwin}}\nn \\[4pt]
 &\quad
   +\frac{\epsilon_0}{2} \int d^3x\,
     \left[
       \frac{( \Pi +\mathcal{P} + F[\vec A])^2}{\epsilon_0^2}
       +  c^2 \vec{B}^2 - \tfrac{c^2}{2} h_{00}\,\vec{B}^2
          + c^2 h_{ij}\,\vec{B}^i \, \vec{B}^j  \right].
\end{align}
\end{widetext}
Here, $\vec B_{\rm tot}$ is the sum of the magnetic field and the gravito-magnetic field.
\subsection{The internal EM fields}
For the internal EM fields (with $h_{0i} \approx 0,  h_{ij} \approx 0)$, we have the action
\begin{align}
    S_{EM}^{\mathrm{Int}} =& \, \frac{\epsilon_0}{4} \int d^3 \vec{x}\; \partial_t^2 (\vec{\mathcal{A}}^{\perp})^2
    -\frac{\epsilon_0}{2} \int d^3 \vec{x}\; \partial_t^2 (\vec{\mathcal{A}}^{\perp})^2 h^{00} \nn\\[4pt]
    &+\frac{1}{2} \int d^3 x\; \big[\vec{j} \cdot \mathcal{A}^{\perp} - \rho \phi\big],
\end{align}
where $\partial_t^2 \left(\vec{\mathcal{A}}^{\perp}\right)^2$ is of order $\order{1/c^4}$, and contribute only at orders beyond the accuracy considered here. To express the internal fields  in terms of the particle position and momentum operators, we evaluate the Maxwell equations.

\section{Maxwell equations on curved spacetime}
The general Maxwell equations on a curved spacetime are given by \cite{Tsagas2005}
\begin{align}
    \nabla_{\nu} F^{\mu \nu} = -\mu_0 J^{\mu}.
\end{align}
Expanding the left-hand side yields,
\begin{align}
    \nabla_{\nu} F^{\mu \nu}
    &= g^{\nu \rho} \nabla_{\nu} \nabla_{\rho} \Anp^{\mu} - g^{\mu \rho} \nabla_{\nu} \nabla_{\rho} \Anp^{\nu},
\end{align}
where $\nabla_{\mu} A_{\nu} = \partial_{\mu} A_{\nu} + \Gamma_{\mu \nu}^{\lambda} A_{\lambda}$ is the covariant derivative, not to be confused with the nabla $\nabla$. After expanding the covariant derivatives, we find
\begin{align}
   \nabla_{\nu} F^{\mu \nu}
   =&\, \Box \Anp^{\mu} - \eta^{\mu \rho} \partial_{0} \partial_{\rho} \Anp^{0}\nn \\[4pt]
   &- h^{\nu \rho} \partial_{\nu} \partial_{\rho} \Anp^{\mu} +h^{\mu \rho} \partial_{0} \partial_{\rho} \Anp^{0}\nn \\[4pt]
   &+ 2 \, \eta^{\nu \rho} \partial_{\rho} \Anp^{\lambda} \Gamma^{\mu}_{\nu \lambda}
   + \Anp^{\lambda} \eta^{\nu \rho} \partial_{\nu} \Gamma^{\mu}_{\rho \lambda} \nonumber \\[4pt]
   &-  2 \, \eta^{\mu \rho} \partial_{\lambda} \Anp^{\nu} \Gamma^{\lambda}_{\nu \rho}
   - \Anp^{\lambda} \eta^{\mu \rho} \partial_{\nu} \Gamma^{\nu}_{\rho \lambda}.
   \label{eq:maxwell-expanded}
\end{align}
At a local inertial frame, the Christoffels evaluated at $\vec r_{\rm I}$ vanish, although expansion around $\vec r_{\rm I}$ reveals curvature effects, demonstrated by the metric (\ref{metric}). We evaluate these effects afterward, and first evaluate the terms containing $\partial \Gamma$ in eq. (\ref{eq:maxwell-expanded}):

\begin{equation}
\mathcal X^\mu
\equiv \Anp^{\lambda}\,\partial^{i}\Gamma^{\mu}{}_{i\lambda}
 - \Anp^{\lambda}\,\eta^{\mu\rho}\,\partial_{i}\Gamma^{i}{}_{\rho\lambda},
\end{equation}
with
\begin{align}
\Gamma^{\mu}_{\nu\lambda}
 = \frac{1}{2}\,\eta^{\mu\alpha}
   ( \partial_{\nu} h_{\lambda \alpha}
    + \partial_{\lambda} h_{\nu \alpha}
    - \partial_{\alpha} h_{\nu\lambda} ),
\end{align}
the second derivatives become $\partial_i\partial_j h_{\alpha\beta}=2 H_{\alpha\beta ij},$
where $h_{\alpha\beta}(t,r)=H_{\alpha\beta ij}(t)\, r^i r^j$, 
and \(H_{\alpha\beta ij}\) given by
\begin{subequations}
\begin{align}
H_{00ij} &\equiv - R_{0 i 0 j},  \\[4pt]
H_{0kij} &\equiv - \tfrac{2}{3} R_{0 i k j},\\[4pt]
H_{klij} &\equiv - \tfrac{1}{3} R_{k i l j}.
\end{align}
\end{subequations}

Thus
\begin{align}
\mathcal X^\mu
 =&\, \Anp^{\lambda}\,\eta^{ij}\,
     \Big\{ \eta^{\mu \alpha}
      ( H_{\alpha \lambda ji} + H_{\alpha i j \lambda} - H_{i \lambda j \alpha} )\nn \\[4pt]
 &- \eta^{\mu \rho}
      ( H_{ \lambda j i \rho} + H_{\rho j i \lambda} - H_{ \rho \lambda ij}) \Big\}.
\end{align}
For $\mu=0$, this reduces to
\begin{equation}
\mathcal X_{0}
   = 2\,R_{00}\,\Anp^{0}
     + \frac{4}{3}\,R_{0m}\,\Anp^{m},
\end{equation}
which vanishes in a vacuum. Similarly, for spatial index $k$,
\begin{equation}
\mathcal X_{k} = 2  \Anp^{\lambda} \left( \delta^{ij} H_{k\lambda ij} - \delta^{ij} H_{\lambda jik}\right),
\end{equation}
which again vanishes after the spatial trace is taken. Hence, the terms containing $\partial \Gamma$ describe coupling to the Ricci vector, which presents curvature generated by the local sources. In a vacuum, the Ricci vector and these terms clearly vanishes. Consequently all terms conainting $\Gamma$ have vanished in the Maxwell equation, and the derivation becomes equivalent to reducing the covariant derivatives to partial derivative. This was similarly done by Schartz and Giuilini in ref.~\cite{SchwartzGiulini2019}, and leads to similar results.

\subsection{Coupling curvature to electric and magnetic fields}\label{curvature-couplings}
 To examine the effects of the external gravitational wave on the internal electrodynamic structure, we may expand $\Gamma$ around $\vec{r}_i$:
\begin{align}
    \Gamma^{\nu}_{\rho \lambda}(\vec{r})
    \simeq \Gamma^{\nu}_{\rho \lambda}(\vec{r}_i)
    + r^k \, \partial_k \Gamma^{\nu}_{\rho \lambda}\big|_{\vec{r} = \vec{r}_i}
    = r^k \, \partial_k \Gamma^{\nu}_{\rho \lambda}\big|_{\vec{r} = \vec{r}_i},
\end{align}
where the first term vanishes by choice of local inertial coordinates. All contributions involving third or higher derivatives of the metric can also be dropped, since the metric perturbation is of order $\mathcal{O}(r^i r^j)$. Inserting this expanded Christoffel symbols into eq. \eqref{eq:maxwell-expanded} one finds
\begin{align}
   \nabla_{\nu} F^{\mu \nu} =& \;\Box \Anp^{\mu}
   - \eta^{\mu \rho}\, \partial_{0} \partial_{\rho} \Anp^{0}
   - h^{\nu \rho}\, \partial_{\nu} \partial_{\rho} \Anp^{\mu}
   + h^{\mu \rho} \,\partial_{0} \partial_{\rho} \Anp^{0} \nonumber \\[4pt]
   & + 2 \eta^{\nu \rho} \, \partial_{\rho} \Anp^{\lambda} \, r^i \partial_i \Gamma^{\mu}_{\nu \lambda}
   - 2 \eta^{\mu \rho} \, \partial_{\lambda} \Anp^{\nu} \, r^i \partial_i \Gamma^{\lambda}_{\rho \nu}.
   \label{eq:expanded-derivativeF}
\end{align}
This shows explicitly how Maxwell’s equations acquire corrections to the electrodynamical fields through derivatives of $\Anp^{\mu}$ interacting with both $h_{\mn}$ and derivatives of the Christoffel symbols, which account for curvature effects.

\subsubsection*{The scalar potential}
We return to the explicit equations for the scalar potential $\phi$ and the transverse vector potential $\vec{\mathcal{A}}^{\perp}$. First the second line in eq.~\eqref{eq:expanded-derivativeF} is evaluated for the temporal component ($\mu = 0$):
\begin{equation}
S \equiv 2 \, \eta^{\nu \rho} \, \partial_\rho \Anp^\lambda \, r^i \partial_i \Gamma^0_{\nu \lambda} 
    - 2 \, \eta^{0 \rho} \, \partial_\lambda \Anp^\nu \, r^i \partial_i \Gamma^\lambda_{\rho \nu}.
\end{equation}
After splitting $\Anp^0$ and $\Anp^p$, and keeping only the terms with two \emph{spatial} derivatives of the metric (i.e., the nonzero terms), all terms involving $\partial_0 \Anp^\mu$ cancel, leaving
\begin{equation}
S = 2 \, r^i \Big[ \partial^m \Anp^0 (\partial_i \Gamma^0_{m0} + \partial_i \Gamma^m_{00}) 
+ \partial^n  \Anp^p (\partial_i \Gamma^0_{np} + \partial_i \Gamma^n_{0p}) \Big].
\end{equation}
With the linearized Christoffel symbols, the first Bianchi identity and the symmetries of the Riemann tensor, the expression finally reduces to
\begin{equation}
S = 4 \, r^i \, R_{0 i 0 m} \, \partial^m \Anp^0 - \frac{4}{3} \, r^i \, \partial^n \Anp^p \, R_{0 p n i}.
\end{equation}
The derivative, $\partial^n \Anp^p $, in the second term can be split into a symmetric and antisymmetric part. Due to the antisymmetry of the Riemann tensor, the symmetric part will cancel, while the antisymmetric part can be written as the field strength $F^{np}$. The second part then becomes a coupling between the quantities $F^{np}$ and $R_{0pni}$, i.e.,
\begin{equation}
S = 4 \, r^i \, R_{0 i 0 m} \, \partial^m \Anp^0- \frac{2}{3} r^i F^{np} R_{0pni}.
\end{equation}
The equation can be rewritten in terms of the electric field $\vec E$ and magnetic field $\vec B$ as well. With $\Anp^0 = \phi/c$ and the electric and magnetic fields given by
\begin{align}
E^m &= - \partial^m \phi = - c \, \partial^m \Anp^0, \nn \\[4pt]
F^{np} &= \partial^n \Anp^p - \partial^p \Anp^n = - \epsilon^{npq} B_q,
\end{align}
the equation becomes
\begin{equation}
S = - \frac{4}{c} \, r^i \, R_{0 i 0 m} \, E^m 
+ \frac{2}{3} \, r^i \, \epsilon^{npq} \, R_{0 p n i} \,  B_q,
\end{equation}
which shows explicitly how $R_{0 i 0 m}$ couples to the electric field, and $R_{0ijk}$ couples to the magnetic field.

The full adjusted Poisson equation becomes
\begin{align}
    \nabla_{\nu} F^{\,0 \nu} =& \, \nabla^2 \vec{\mathcal{A}}^0 - \frac{1}{c}  h^{i 0}\, \partial_i \partial_{t}  \vec{\mathcal{A}}^0 - h^{i j} \, \partial_i \partial_j \vec{\mathcal{A}}^0 \nonumber \\
&+4 \, r^i \, R_{0 i 0 n} \, \partial^n \Anp^0 - \frac{4}{3} \, r^i \, \partial^n \Anp^j \, R_{0 j n i} \nonumber \\
=& - \mu_0 j^{0}\; ( 1 + \frac{1}{2} h_{00}).
\end{align}
After converting $\vec{\mathcal{A}}^0 = -(1+ h^{00}) \, \vec{\mathcal{A}}_0$, with $\vec{\mathcal{A}}_0 =- \phi/c$, and confining to $h_{00} \propto R_{0i0j}$ contributions, the field equation reads
\begin{align}
    \nabla^2 \phi+ 4 \, r^i \, R_{0 i 0 n} \, \partial^n \phi
    &= - \frac{\rho}{\epsilon_0}\left(1 -\tfrac{1}{2} h_{00}\right).
    \label{eq:phi-eq}
\end{align}
The appearance of $h_{00}$ in the current density $\rho$ illustrates how a gravitational wave perturbs charge and current densities that generate the electromagnetic field. \\

Introducing $h_{\mu\nu} = \epsilon h_{\mu\nu}$ with  $\epsilon \ll 1$, and $\phi = \phi_{(0)} + \epsilon \phi_{(1)} + \mathcal{O}(\epsilon^2),$ the Maxwell’s equations can be solved order by order. At zeroth order, one recovers the Poisson equation in flat spacetime \cite{sonnleitner2018, Jackson1998},
\begin{align}
    \nabla^2 \phi_{(0)} &= -\frac{\rho}{\epsilon_0}, 
\end{align}
with solution
\begin{align}
    \phi_{(0)} &= \sum_{\mathrm{I}} \frac{e_{\mathrm{I}}}{4 \pi \epsilon_0 |\vec{r}-\vec{r}_{\mathrm{I}}|}.
\end{align}
At first order in $\epsilon$ we have
\begin{align}
    \nabla^2 \phi_{(1)}+ 4 \, r^i \, R_{0 i 0 n} \, \partial^n \phi_{(0)}
    &= \frac{\rho}{2 \epsilon_0} h_{00}.
    \label{eq:phi-eq}
\end{align}
Integrating the differential equation yields
\begin{align}
    \phi_{(1)}  =& - \frac{e_{\mathrm{I}} R_{0i0n}}{\pi \epsilon_0}\left(\frac{\delta^{in}}{|\vec r-\vec r_{\mathrm{I}}|}- \frac{(\vec r-\vec r_{\mathrm{I}})^n(\vec r-\vec r_{\mathrm{I}})^j}{|\vec r-\vec r_{\mathrm{I}}|^3}\right) \nn \\
    &+  \frac{e_{\mathrm{I}}}{8 \pi \epsilon_0}  \frac{h_{00}}{| \vec{r}-\vec{r}_{\mathrm{I}}|},
\end{align}
where the first term vanishes in vacuum $\delta^{in} R_{0i0n} = R_{00} =0$.

\subsubsection*{The vector potential}
For the spatial components ($\mu = k$), the field equations become more involved, although they are derived by the same procedure outlined above. The evaluation of
\begin{align}
S^{k} \;=\; 2\,\eta^{\nu\rho}\,\partial_\rho \Anp^\lambda \; r^i \partial_i \Gamma^k_{\nu\lambda}
-2\,\eta^{k\rho}\,\partial_\lambda \Anp^\nu \; r^i \partial_i \Gamma^\lambda_{\rho\nu},
\end{align}
gives
\begin{align}\label{eq:Ek_partialh}
S^{k} =&\; 2\,r^i\Big[
\partial_0 \Anp^0\,\partial_i\partial_k h_{00}
+\partial_0 \Anp^p\,\partial_i\partial_k h_{0p} \nn \\[4pt]
&+\partial^n \Anp^0\big(\partial_i\partial_n h_{0k}-\partial_i\partial_k h_{0n}\big) \nn \\[4pt]
&+\partial^n \Anp^p\big(\partial_i\partial_n h_{kp}-\partial_i\partial_k h_{np}\big)
\Big].
\end{align}
Substituting the metric components and using the algebraic symmetries of the Riemann tensor and the first Bianchi identity, we arrive at
\begin{align}
S^{k} =& -4\,r^i\,R_{0 i 0 k}\,\partial_0 \Anp^0
+ \frac{4}{3}\,r^i\,R_{0 p k i}\,\partial_0 \Anp^p \\
&+ \frac{8}{3}\,r^i\,\partial^n \Anp^0 \, R_{0 n k i}
+\frac{8}{3}\,r^i\,\partial^n \Anp^p\Big( R_{k n p i} + 2 R_{k p i n} \Big). \nn
\end{align}
Here, different index ordering may result in a different sign. 

Once again, we may split \(\partial^n \Anp^p\) into
\begin{align}
\partial^n \Anp^p = \tfrac12(\partial^n \Anp^p + \partial^p \Anp^n) + \tfrac12 F^{np},
\end{align}
and drop the symmetric part wherever the Riemann factor is antisymmetric in \(n,p\). Only the antisymmetric part of \( \partial^n \Anp^p \) contributes, i.e. the spatial tensor \(F^{np} = \partial^n \Anp^p - \partial^p \Anp^n\). 
Alternatively, the equation can be reorganized into pieces proportional to the electric field, \(\partial^n \Anp^0 = - \tfrac{1}{c} E^n\), and pieces proportional to the magnetic field via \(F_{np} = -\epsilon_{npq} B^q\), as was done for the $\mu=0$ equation.

Expanding the full expression for $\mu = k$ yields
\begin{align}
\nabla_{\nu} F^{k \nu} 
=&\, \Box \Anp^k - \frac{1}{c} \, \partial_t \partial^k \vec{\mathcal{A}}^0 - h^{\nu \rho}\, \partial_{\nu} \partial_{\rho} \Anp^k + \frac{1}{c}h^{k\rho}\, \partial_t \partial_{\rho} \vec{\mathcal{A}}^0\nonumber \\
&-4\,r^i\,R_{0 i 0 k}\,\partial_0 \Anp^0
+ \frac{4}{3}\,r^i\,R_{0 p k i}\,\partial_0 \Anp^p \nn \\
&+ \frac{8}{3}\,r^i\,\partial^n \Anp^0 \, R_{0 n k i} 
+\frac{8}{3}\,r^i\,\partial^n \Anp^p\Big( R_{k n p i} + 2 R_{k p i n} \Big).
\end{align}
which, after retaining $h_{00}$ contributions, reduces to 
\begin{align}
   \nabla_{\nu} F^{k \nu}
   =&\, \Box \Anp^k - \frac{1}{c}\partial_t \partial^k \vec{\mathcal{A}}^0 - \frac{h^{00}}{c^2} \partial_t^2  \Anp^k - 4\,r^i\,R_{0 i 0 k}\,\partial_0 \Anp^0.
\end{align}
Notice that $\Anp$ is of order $\order{1/ c^2}$, making $ \frac{h^{00}}{c^2}\partial_t^2  \Anp^k$ of order $\order{1 / c^4}$. Thus, the vector potential satisfies
\begin{align}
   \Box \Anp_k - \frac{1}{c^2}\partial_t \partial_k \phi - \frac{4}{c^2} \,r^i\,R_{0 i 0 k}\,\partial_t \phi
   &= - \mu_0 j^i \delta_{ik} \left(1 - \tfrac{1}{2} h_{00}\right).
   \label{eq:Ai-eq}
\end{align}
Likewise, the vector potential acquires corrections proportional to the metric perturbation $h_{00}$, which modify the sources $\vec{j}$, as well as corrections from $R_{0i0k}$ affecting the propagation of the field itself.

We similarly introduce $\mathcal{A} = \mathcal{A}_{(0)} + \epsilon \mathcal{A}_{(1)} + \mathcal{O}(\epsilon^2).$
At zeroth order we have \cite{sonnleitner2018, Jackson1998},
\begin{align}
    \left(\nabla^2 - \frac{1}{c^2}\partial_t^2\right)\mathcal{A}_{(0)} &= - \mu_0 \left( \vec{j} - \epsilon_0 \nabla \partial_t \phi \right),
\end{align}
with solution
\begin{align}
    \mathcal{A}_{(0)} &= \sum_{\mathrm{I}} \frac{ e_{\mathrm{I}}}{8\pi \epsilon_0 c^2} 
    \left( \frac{\vec{v}_{\mathrm{I}}}{ |\vec{r}-\vec{r}_{\mathrm{I}}|} + \frac{(\vec{r} - \vec{r}_{\mathrm{I}}) [ \vec{v}_{\mathrm{I}} \cdot (\vec{r} - \vec{r}_{\mathrm{I}})]}{|\vec{r} - \vec{r}_{\mathrm{I}}|^3}\right).
\end{align}
For solving the first order correction of the vector potential, notice that the continuity equation allows us to rewrite $\partial_t \rho = - \nabla \cdot \vec j$. Integrating yields
\begin{align}
\Anp_{(1)}
=& \sum_k \frac{R_{0i0k}}{ \pi \epsilon_0 c^2} \left(\frac{v^i_{\mathrm{I}}}{|\vec r-\vec r_{\mathrm{I}}|}
-\frac{(\vec r-\vec r_{\mathrm{I}})^i \;\vec v \cdot\vec (\vec r-\vec r_{\mathrm{I}})}{|\vec r-\vec r_{\mathrm{I}}|^3}\right)\nn \\ &+ \frac{\mu_0 e_{\mathrm{I}} h_{00}}{16 \pi } \frac{\vec v_{\mathrm{I}}}{|\vec r - \vec r_{\mathrm{I}}|}. 
\end{align}
To summarize, the full coupling between source and internal fields, interacting with gravitational waves, is given by
\begin{align}
    &\frac{1}{2}\,  (\vec j \cdot \vec{\mathcal{A}} - \rho \, \phi) 
    =\nn \\[4pt]
    & - \sum_{\mathrm{I}, \mathrm{J}}  \frac{e_{\mathrm{I}} e_{\mathrm{J}} }{4 \pi \epsilon_0 r} \left(1-  \tfrac{1}{2} h_{00}\right)  - \frac{e_{\mathrm{I}} e_{\mathrm{J}} R_{0i0n}}{\pi \epsilon_0}  \frac{r^n r^j}{r^3}\nn \\[4pt]
    & + \frac{e_{\mathrm{I}} e_{\mathrm{J}}}{8\pi \epsilon_0 c^2} 
    \left(\frac{\vec{v}_{\mathrm{I}} \cdot \vec{v}_{\mathrm{J}}}{ r} \left(1-  \tfrac{1}{2} h_{00}\right) + \frac{(\vec{v}_{\mathrm{I}} \cdot \vec{r}) (\vec{v}_{\mathrm{J}} \cdot \vec{r})}{r^3}\right) \nn \\[4pt]
    & + \sum_k \frac{R_{0i0k}}{ \pi \epsilon_0 c^2} \left(\frac{v^i_{\mathrm{I}} v^k_{\mathrm{J}}}{r}
    -\frac{ v_{\mathrm{J}}^k r^i\; (\vec v_{\mathrm{I}} \cdot \vec r) + \mathrm{I} \leftrightarrow \mathrm{J}}{2 r^3} \right),
\end{align}
where $h_{00}$ can be written as the gravitationally induced quadrupole term $h_{00} = R_{0i0n}  r^i r^n$.
The first line reproduces the Coulomb interaction between charges, corrected by this gravitationally induced quadrupole on the charges and corrections to the scalar field from curvature. The remaining part corresponds to the magneto-static (Darwin) interaction, also dressed by similar gravitational corrections.

\section{The full Hamiltonian}

We collect all obtained results for the atom-light-gravity system and summarize the leading contributions to the full Hamiltonian. Starting from the kinetic Hamiltonian,

\begin{align}
H_{\text{kin}}
=& \sum_{\mathrm{I}} \left[
\frac{(\vec{p}_{\mathrm{I}} 
    + \frac{1}{2}\, \vec{d} \times \vec{B}(\vec{R}))^{\,2}}{2 \tilde{m}_{\mathrm{I}}}
- \frac{\vec p_{\mathrm{I}}^{\,4}}{8 \tilde{m}_{\mathrm{I}}^3 c^2}
\right],
\end{align}
where $\tilde{m} = m + \delta m$. The PWZ transformation has been implemented, and $\vec B$ can be taken as the sum of the magnetic field and the gravito-magnetic field. The interaction part is

\begin{align}
&H_{\text{int\,EM}}
= \;
\sum_{\mathrm{I}, \mathrm{J}}  
\frac{e_{\mathrm{I}} e_{\mathrm{J}} }{4 \pi \epsilon_0 r} \left(1-  \tfrac{1}{2} h_{00}\right)  
+ \frac{e_{\mathrm{I}} e_{\mathrm{J}} R_{0i0n}}{\pi \epsilon_0}  \frac{r^n r^j}{r^3} \nonumber \\[4pt]
&+ \frac{e_{\mathrm{I}} e_{\mathrm{J}}}{8\pi \epsilon_0 c^2 \tilde{m}_{\mathrm{I}} \tilde{m}_{\mathrm{J}}} 
\left(
\frac{\vec{p}_{\mathrm{I}} \cdot \vec{p}_{\mathrm{J}}}{ r} \left(1-  \tfrac{1}{2} h_{00}\right)
+ \frac{(\vec{p}_{\mathrm{I}} \cdot \vec{r}) (\vec{p}_{\mathrm{J}} \cdot \vec{r})}{r^3}
\right)\nonumber\\[4pt]
&+
\sum_k 
\frac{R_{0i0k}}{ \pi \epsilon_0 c^2 \tilde{m}_{\mathrm{I}} \tilde{m}_{\mathrm{J}}}
\left(
\frac{p^i_{\mathrm{I}}\, p^k_{\mathrm{J}}}{r}
-\frac{ p_{\mathrm{J}}^k\,  r^i\; (\vec p_{\mathrm{I}} \cdot \vec r) + \mathrm{I} \leftrightarrow \mathrm{J}}{2 r^3} 
\right).
\end{align}
The external field Hamiltonian is the standard one with coupling to the $h_{00}$ and $h_{ij}$ metric components, as well as a momentum shift:

\begin{align}
H_{\text{ext\,EM}}
= \int d^3x\,
\left[
\frac{\bar{\Pi}^2}{2\epsilon_0}
+ \frac{\epsilon_0 c^2}{2}
\left( 
\vec B^2 
- \tfrac{1}{2} h_{00}\,\vec{B}^2
+ h_{ij}\,\vec{B}^i \, \vec{B}^j
\right)
\right],
\end{align}
where $\Bar{\Pi} = \Pi + \mathcal{P} + F[A]$.

% ==================================================
% QUANTUM HAMILTONIAN
% ==================================================

\subsection{Quantum Hamiltonian}

For the quantum Hamiltonian, we adopt the operator ordering prescribed in Ref.~\cite{sonnleitner2018}, which ensures a consistent quantum treatment of the electromagnetic interaction in curved spacetime. Writing the Riemann tensor schematically as $R(t)$, the interaction term $\vec{j} \cdot \Anp_{(1)}$ can be expressed as
\begin{align}
 \vec{j} \cdot \Anp_{(1)} \propto & \; 
\frac{\vec{p}_{\mathrm{I}} \cdot \vec{p}_{\mathrm{J}}}{r} 
\, R(t)\cdot \vec{r}^2  + \frac{  R(t)\cdot(\vec p_{\mathrm{I}} \vec p_{\mathrm{J}})}{r}\nn \\[4pt]
&-\frac{(\vec{p}_{\mathrm{I}} \cdot \vec{r})\; R(t) \cdot (\vec{p}_{\mathrm{J}} \vec{r})}{r^3}  
+ \mathrm{I} \leftrightarrow \mathrm{J}\nonumber\\[4pt]
=&\; R_{0i0j}\, r^i 
\left(\vec{p}_{\mathrm{I}} \cdot \frac{1}{r} \vec{p}_{\mathrm{J}}\right) 
r^j + R_{0i0j} \left(\vec{p}_{\mathrm{I}}^i \cdot \frac{1}{r} \vec{p}_{\mathrm{J}}^j\right) \nn \\[4pt]
&- (\vec{p}_{\mathrm{I}} \cdot \vec{r}) \frac{1}{r^3}
\left(R_{0i0j} \, \vec{p}_{\mathrm{J}}^i r^j\right)+ \mathrm{I} \leftrightarrow \mathrm{J}.
\end{align}
With this chosen operator ordering, $H_{\text{int EM}}$ becomes
\begin{align}
H_{\text{int EM}}
=&
\sum_{\mathrm{I}, \mathrm{J}}  
\frac{e_{\mathrm{I}} e_{\mathrm{J}} }{4 \pi \epsilon_0 r} \left(1-  \tfrac{1}{2} h_{00}\right)
+ \frac{e_{\mathrm{I}} e_{\mathrm{J}} R_{0i0n}}{\pi \epsilon_0}  \frac{r^n r^j}{r^3}\nonumber\\[4pt]
&+
\frac{e_{\mathrm{I}} e_{\mathrm{J}}}{16\pi \epsilon_0 c^2 m_{\mathrm{I}} m_{\mathrm{J}}}  
\left(
\vec{p}_{\mathrm{I}} \cdot \tfrac{1}{r} \vec{p}_{\mathrm{J}} 
+ (\vec{p}_{\mathrm{I}} \cdot \vec{r}) \tfrac{1}{r^3} (\vec{p}_{\mathrm{J}} \cdot \vec{r})
\right)\nonumber\\[8pt]
&+
\frac{e_{\mathrm{I}} e_{\mathrm{J}} R_{0i0j}}{32\pi \epsilon_0 c^2 m_{\mathrm{I}} m_{\mathrm{J}}}  
\, r^i 
\left(\vec{p}_{\mathrm{I}} \cdot \tfrac{1}{r} \vec{p}_{\mathrm{J}}  \right) 
r^j\nonumber\\[8pt]
&+
\frac{e_{\mathrm{I}} e_{\mathrm{J}}}{2 \pi^2 \epsilon_0 c^2 m_{\mathrm{I}} m_{\mathrm{J}}}
\Big[
R_{0i0j} \left(\vec{p}_{\mathrm{I}}^i \cdot \tfrac{1}{r} \vec{p}_{\mathrm{J}}^j\right) \nonumber\\[8pt]
&- (\vec{p}_{\mathrm{I}} \cdot \vec{r}) \tfrac{1}{r^3}
\left(R_{0i0j} \, \vec{p}_{\mathrm{J}}^i r^j \right)
\Big]
+ \mathrm{I} \leftrightarrow \mathrm{J}.
\end{align}

% ==================================================
% COM AND RELATIVE COORDINATES
% ==================================================

\subsection{Separation into center-of-mass and relative coordinates}

Lastly, the Hamiltonian is expressed in center-of-mass (COM) and relative coordinates. Define
\begin{align}
& \vec{P} = \vec{p}_1 + \vec{p}_2, 
&& \vec{R} = \frac{\tilde{m}_1 \vec{r}_1 + \tilde{m}_2 \vec{r}_2}{\tilde{M}}, 
\nonumber\\
& \vec{p}_{1,2} = \frac{\tilde{m}_{1,2}}{\tilde{M}}\,  \vec{P} \pm \vec{p}_r, 
&& \vec{r}= \vec{r}_1 - \vec{r}_2,
\end{align}
with total mass $\tilde{M} = \tilde{m}_1 + \tilde{m}_2$ and reduced mass $\mu = \tilde{m}_1 \tilde{m}_2 / \tilde{M}$. In these variables, the Hamiltonian separates into five contributions:
\begin{align}
H_C = &\, \frac{\vec P^2}{2 M} \Big[1 - \frac{\vec{P}^2}{4 M^2 c^2}  \nn \\[4pt]
    &- \frac{1}{M c^2}\Big( \,\frac{\vec{p}_r^2}{2\mu}
    - \frac{e_1 e_2}{4 \pi \epsilon_0 r}\left(1+\tfrac{1}{2} R_{0k0l} \vec{r}^k \vec{r}^l \right) \Big) \Big]\nn \\[4pt]
    &+ \frac{1}{M^2 c^2} \frac{e_1 e_2}{\pi \epsilon_0}\left(\frac{P^i P^j \, R_{0i0j}}{ r} \right),  \\[0.75cm]
H_A =&\, \frac{\vec{p}_r^2}{2 \mu}\left[1 - \frac{\vec{p}_r^2}{4\mu^2 c^2}\frac{m_1^3 +  m_2^3}{M^3}\right] \nn \\[4pt] 
    &- \frac{e_1 e_2}{4 \pi \epsilon_0 r}\left(1 + \tfrac{1}{2} R_{0i0j} \vec{r}^i \vec{r}^j \right) 
    - \frac{e_1 e_2 R_{0i0j} r^i r^j}{\pi \epsilon_0 r^3}\nn  \\[4pt]
    &- \frac{e_1 e_2}{8 \pi \epsilon_0 c^2 \mu M} \left(\vec{p}_r \cdot \tfrac{1}{r} \vec{p}_r 
    + \tfrac{1}{2} R_{0i0j}\, \vec{r}^i \left(\vec{p}_r \cdot \tfrac{1}{r} \vec{p}_r \right) \vec{r}^j \right)\nn  \\[4pt]
    &- \frac{e_1 e_2}{8 \pi \epsilon_0 c^2 \mu M} \left( (\vec{p}_r \cdot \vec{r}) \tfrac{1}{r^3}(\vec{p}_r \cdot \vec{r}) \right) \nn  \\[4pt]
    & -\frac{e_1 e_2}{\pi \epsilon_0 c^2 \mu M} \left( R_{0i0j}\; \vec{p_r}^i \, \tfrac{1}{r}\vec{p_r}^j- (\vec{p}_r \cdot \tfrac{1}{r^3}\vec{r}) (R_{0i0j} \vec{p}_r^i \vec{r}^j) \right),  \\[0.75cm]
H_\chi =& - \frac{(\vec{P}\cdot \vec{p}_r)^2}{2 \mu M^2 c^2} 
    - \frac{m_2-m_1}{2 M^2 \mu^2 c^2} (\vec{P}\cdot \vec{p}_r) \vec{p}_r^2\nn   \\[4pt]
    &+ \frac{e_1 e_2}{8 \pi \epsilon_0 c^2 M^2 r^3} \left((\vec{P}\cdot \vec{r})^2
    - 8 (\vec{P}\cdot \vec{r})(R_{0i0j}\vec{P}^i \vec{r}^j) \right)\nn  \\[4pt]
    &+ \frac{e_1 e_2 (m_2-m_1)}{16 \pi \epsilon_0 c^2 \mu M^2} \Big(\vec{P}\cdot \tfrac{1}{r}\vec{p}_r 
    + \tfrac{1}{2} R_{0i0j}\, \vec{r}^i \left(\vec{P}\cdot \tfrac{1}{r} \vec{p}_r\right) \vec{r}^j \nn \\[4pt]
    &+ 16 R_{0i0j}\,P^i \; \tfrac{1}{r} p_r^j\Big) \nn  \\[4pt]
    &+ \frac{e_1 e_2 (m_2-m_1)}{16 \pi \epsilon_0 c^2 \mu M^2} \left( (\vec{P}\cdot \vec{r}) \tfrac{1}{r^3} (\vec{p}_r \cdot \vec{r}) \right)\nn  \\[4pt]
    &- \frac{e_1 e_2 (m_2-m_1)}{2 \pi \epsilon_0 c^2 \mu M^2} (\vec{P} \cdot \vec{r}) \tfrac{1}{r^3} (R_{0i0j} \vec{p}_r^i \vec{r}^j) + \text{h.c.},
    \end{align}
    \begin{align}
H_L =& \,\frac{\epsilon_0}{2} \int d^3 r \left((\vec{E}_{\perp}+ \frac{1}{\epsilon_0} F[\vec B])^2 +  c^2 \; \vec B^2_{\perp} \right) \nn \\[4pt]
    &+\frac{\epsilon_0}{2} \int d^3 r \Big( (\vec{E}_{\perp}^2 -  c^2 \vec{B}^2_{\perp})\tfrac{1}{2} h_{00} \nn \\[4pt]
    &-(\vec{E}_i^{\perp} \, \vec{E}_j^{\perp} - c^2 \vec{B}_i^{\perp} \, \vec{B}_j^{\perp}) h^{ij} \Big), \\[0.75cm]
H_{AL} =& - \vec d \cdot \vec E^{\perp} (\vec R) + \frac{1}{2M} ( \vec P \cdot [\vec d \times\vec B(\vec R)])\nn \\[4pt]
    &+ \frac{m_2-m_1}{4 m_1 m_2} \left(\vec{p}_r \cdot [\vec d \times\vec B(\vec R)] + \text{h.c.} \right) \nn  \\[4pt]
    &+ \frac{1}{8\mu} (\vec d \times \vec B(\vec R))^2 + \frac{1}{2 \epsilon_0} \int d^3 \vec x \; (\mathcal{P}^{\perp})^2.
\end{align}

\section{Discussion}

The atomic Hamiltonian receives two distinct gravitational contributions. The first originates from the metrics influence on the source, producing the tidal correction that modifies the binding potential through variations in particle separation. In the transverse-traceless (TT) gauge, $R_{0i0j}^{\mathrm{TT}}=\tfrac{1}{2c^{2}}\ddot h^{\mathrm{TT}}_{ij}$, which implies that the corresponding atomic term scales as $\mathcal{O}(1/c^{2})$, whereas the analogous term in the center-of-mass (COM) Hamiltonian scales as $\mathcal{O}(1/c^{4})$, i.e., below our energy cut-off. This hierarchy reflects the relation $H_{C}\simeq H_{A}/(M c^{2})$, however, which suppresses COM contributions by an additional factor of $1/c^{2}$. If the expansion in $1/c^2$ is taken to infinity, all internal corrections eventually contribute to $H_{C}$.

Explicitly, for the first gravitational contribution we find
\[
H_{C}\propto 
\frac{\vec P^{\,2}}{2M}
\frac{1}{M c^{2}}
\left(
 \frac{e_{1} e_{2}}{8\pi\epsilon_{0} r}
 R_{0k0l}^{\mathrm{TT}} r^{k} r^{l}
\right),
\]
and
\[ 
H_{A}\propto
\frac{e_{1} e_{2}}{8\pi\epsilon_{0} r}
R_{0k0l}^{\mathrm{TT}} r^{k} r^{l},
\]
which is consistent with the expected scaling. This behavior, however, arises because $h_{0i}$ and $h_{ij}$ were omitted in computing the Coulomb potential and Darwin term; including them would introduce deviations, as demonstrated by Schwartz and Giulini \cite{SchwartzGiulini2019}. Since distances and momentum norms depend on the metric, the decomposition of kinetic and interaction energy inherits a metric dependence, resulting in an ambiguity between inertial and gravitational mass. This ambiguity disappears when all quantities are expressed as local quantities by multiplying with the physical spatial metric $^{(3)}g$.

The second gravitational contribution follows from curvature coupling to the fields. Two qualitatively different structures appear. The internal Hamiltonian contains a term,
\[
-\frac{e_{1}e_{2}}{\pi\epsilon_{0}}\,\frac{R_{0i0j}\,r^{i} r^{j}}{r^{3}},
\]
arising from corrections to spatial derivatives of the scalar field. By contrast, the COM Hamiltonian acquires a curvature-momentum coupling,
\[
\frac{1}{M^{2}c^{2}}
\,\frac{e_{1}e_{2}}{\pi\epsilon_{0}}
\frac{P^{i}P^{j}}{r}\,R_{0i0j},
\]
originating from corrections to the time-varying scalar field (or spatial derivative of the vector field). This term does not renormalize the inertial mass and instead modifies the COM kinetic energy. Similarly, the internal $1/r^{3}$ correction has no counterpart in $H_{C}$, so one cannot write $H_{C}\sim H_{A}/(M c^{2})$. Instead,
\[
H_{C}=
\frac{\vec P^{\,2}}{2M}
\!\left[
 1-\frac{H_{A}}{M c^{2}}
\right]
+ \frac{e^{2}P^{i}P^{j}R_{0i0j} r^{2}}{M^{2}c^{2}\pi\epsilon_{0} r^{3}}
- \frac{e^{2}\vec P^{\,2} R_{0i0j} r^{i} r^{j}}{2M^{2}c^{2}\pi\epsilon_{0} r^{3}},
\]
showing that variations in internal energy can induce a genuine acceleration. These contributions naturally appear as cross terms involving both $\vec P$ and $\vec r$, and may be grouped with the other $1/r^{3}$ cross terms collected in $H_{\chi}$. However, the mathematical origin of these terms suggests they belong in $H_C$ and $H_A$.

Transforming $\vec r$ and $\vec P$ with respect to the physical metric, as discussed above, does not modify their contraction with $R_{0i0j}$. Furthermore, the $R$-dependent terms are of order $\mathcal{O}(h)$, hence to linear order, the operators are transformed with the Minkowski metric, leaving them unchanged. A complete treatment of metric effects requires going beyond the linear approximation.

\subsubsection{Comment on equivalence principle}
The shift of an atom’s binding energy reflects a change in gravitational mass equal to its change in inertial mass, consistent with the weak equivalence principle. This principle holds when the gravitational field is \emph{effectively uniform} over the system, so gravity couples only to the total energy and is insensitive to internal structure. When spacetime curvature varies significantly across the system, however,  the assumptions of the weak equivalence principle are no longer satisfied. In other words, the weak equivalence principle applies to effectively point-like bodies, where gravity couples to the total energy of the system but cannot resolve its internal structure, and tidal effects are negligible. In setup presented here, the equivalence principle does not apply.

\section{Summary}

We have extended the analyses of Sonnleitner and Barnett, as well as those of Schwartz and Giulini, to the case of atomic systems interacting with dynamical gravitational waves. Earlier work investigated mass-energy equivalence in flat spacetime or post-Newtonian corrections in static gravitational fields; here, we incorporate curvature-dependent interactions in fully dynamical spacetimes.

The derived Hamiltonian contains two forms of curvature coupling. First, an internal quadrupole term from the coupling between the Riemann tensor and the spatial derivatives of the scalar potential. Second, a COM-momentum coupling arising from the coupling of the Riemann tensor to the vector potential. At first glance, these contributions demonstrate that not all internal structural changes correspond to a simple rescaling of the inertial mass. Instead, curvature couplings can generate corrections to the momentum arising from acceleration.

Finally, gravitational waves also couple directly to the source, producing the familiar quadrupole tidal interaction. If defined with respect to the physical metric, this term is consistent with post-Newtonian results. It contributes to the total mass via mass-energy equivalence.

\bibliography{library}

@article{Abbott2016,
  author       = {LIGO Scientific and Virgo Collaboration},
  title        = {Observation of Gravitational Waves from a Binary Black Hole Merger},
  journal      = {Physical Review Letters},
  volume       = {116},
  pages        = {061102},
  year         = {2016},
  doi          = {10.1103/PhysRevLett.116.061102}
}

@article{LIGO2009,
  author       = {LIGO Scientific Collaboration},
  title        = {LIGO: The Laser Interferometer Gravitational-Wave Observatory},
  journal      = {Reports on Progress in Physics},
  volume       = {72},
  pages        = {076901},
  year         = {2009},
  doi          = {10.1088/0034-4885/72/7/076901}
}

@misc{Virgo2021,
  author       = {Virgo Collaboration},
  title        = {Virgo - Observing the Universe at the Frontier of Knowledge},
  howpublished = {\url{https://www.virgo-gw.eu/}},
  year         = {2021},
  note         = {Accessed: 2025-11-04}
}

@misc{KAGRA2025,
  author       = {National Astronomical Observatory of Japan},
  title        = {KAGRA Project - Research},
  howpublished = {\url{https://gwpo.nao.ac.jp/en/research/kagra.html}},
  year         = {2025},
  note         = {Accessed: 2025-11-04}
}

@article{Akutsu2019,
  author       = {KAGRA collaboration},
  title        = {KAGRA: 2.5 generation interferometric gravitational-wave detector},
  journal      = {Nature Astronomy},
  volume       = {3},
  pages        = {35-40},
  year         = {2019},
  doi          = {10.1038/s41550-018-0658-y},
}

@misc{LISA2025,
  author       = {European Space Agency},
  title        = {LISA - First Gravitational-Wave Detector in Space},
  howpublished = {\url{https://www.esa.int/Science_Exploration/Space_Science/LISA}},
  note         = {Accessed: 2025-11-04},
  year         = {2025}
}

@misc{ESASpaceMicroscopeUFF,
  author       = {European Space Agency (ESA)},
  title        = {Space Microscope to test universality of freefall},
 url = {https://www.esa.int/Science_Exploration/Space_Science/Space_Microscope_to_test_universality_of_freefall},
  note         = {Accessed: 2025-10-31},
  year         = {2016}
}

@misc{ZARMDropTowerBremen,
  author       = {ZARM -- Zentrum für angewandte Raumfahrttechnologie und Mikrogravitation},
  title        = {Phenomena in Weightlessness – What Happens in the Bremen Drop Tower?},
url = {https://www.zarm.uni-bremen.de/en/in-focus/phenomena-in-weightlessness},
  note         = {Accessed: 2025-10-31},
  year         = {2025}
}

@article{Babiker2024,
  author       = {M. Babiker},
  title        = {Gauge and unitary transformations in multipolar quantum optics},
  journal      = {Philosophical Transactions A: Mathematical, Physical and Engineering Sciences},
  volume       = {382},
pages={20230330},
  year         = {2024},
  doi          = {10.1098/rsta.2023.0330},
  pmcid        = {PMC11667583},
  url          = {https://pmc.ncbi.nlm.nih.gov/articles/PMC11667583/},
 note = {PubMed: \url{https://pmc.ncbi.nlm.nih.gov/articles/PMC11667583/}}
}

@book{Jackson1998,
    author = {J. D. Jackson},
    title = {Classical Electrodynamics},
    publisher = {John Wiley \& Sons, New York},
    year = {1962},
url={https://archive.org/details/john-david-jackson.-classical-electrodynamics}
}

@book{maggiore,
title = {Gravitational Waves Volume 1: Theory and Experiments},
author = {Maggiore, M.},
publisher = {Oxford university press},
isbn = {978-0-19-857074-5},
year = {2008},
}

@book{carroll2024,
    author = {Carroll, S},
    title = {Spacetime and Geometry. An introduction to general relativity},
    isbn = {978-1-292-02663-3},
    publisher = {Pearson Education Limited},
    year = {2014}
}

@book{Misner1973,
    author = {C. W. Misner and S. T. Kip and J. A. Wheeler},
    title = {Gravitation},
    publisher = {W. H. Freeman, San Francisco},
    year = 1973
}

@article{sonnleitner2018,
  title = {Mass-Energy and Anomalous Friction in Quantum Optics},
  author = {Sonnleitner, M. and Barnett, S. M.},
  year = {2018},
  journal = {Physical Review A},
  volume = {98},
  pages = {042106},
  publisher = {American Physical Society},
  doi = {10.1103/PhysRevA.98.042106}
}

@article{SonnleitnerTrautmann2017,
  title = {Will a Decaying Atom Feel a Friction Force?},
  author = {Sonnleitner, M. and Trautmann, N. and Barnett, S. M.},
  journal = {Physical Review Letters},
  volume = {118},
  pages = {053601},
  year = {2017},
  publisher = {American Physical Society},
  doi = {10.1103/PhysRevLett.118.053601}
}

@article{BarnettSonnleitner2018,
  author       = {Barnett, S. M. and Sonnleitner, M.},
  title        = {Vacuum friction},
  journal      = {Journal of Modern Optics},
  volume       = {65},
  pages        = {23--29},
  year         = {2018},
  doi          = {http://doi.org/10.1080/09500340.2017.1374482}
}

@article{SchwartzGiulini2019,
  title = {Post-Newtonian Hamiltonian description of an atom in a weak gravitational field},
  author = {Schwartz, P. K. and Giulini, D.},
  journal = {Physical Review A},
  volume = {100},
  pages = {052116},
  year = {2019},
  publisher = {American Physical Society},
  doi = {10.1103/PhysRevA.100.052116}
}

@article{schwartz2020,
  title = {Post-Newtonian description of quantum systems in gravitational fields},
  author = {Schwartz, P. K.},
  year = {2020},
  archiveprefix = {arXiv},
  eprint = {2009.11319},
}

@article{Zych2019,
  title = {Gravitational mass of composite systems},
  author = {Zych, M. and Rudnicki, \L{}. and Pikovski, I.},
  journal = {Physical Review D},
  volume = {99},
  pages = {104029},
  year = {2019},
  publisher = {American Physical Society},
  doi = {10.1103/PhysRevD.99.104029},
}

@article{Zych2018,
  title={Quantum formulation of the equivalence principle},
  author={Zych, M. and Brukner, {\v{C}}.},
  journal={Nature Physics},
  volume={14},
  pages={1027-1031},
  year={2018},
  publisher={Nature Publishing Group},
doi={https://doi.org/10.1038/s41567-018-0197-6}
}

@article{Amelino1999,
  author       = {G. Amelino-Camelia},
  title        = {Gravity-wave interferometers as quantum-gravity detectors},
  journal      = {Nature},
  volume       = {398},
  pages        = {216-218},
  year         = {1999},
  doi          = {10.1038/18377}
}

@article{Hogan2012,
  author       = {C. J. Hogan},
  title        = {Interferometers as probes of Planckian quantum geometry},
  journal      = {Phys. Rev. D},
  volume       = {85},
  pages        = {064007},
  year         = {2012},
  doi          = {10.1103/PhysRevD.85.064007}
}

@article{Tsagas2005,
year = {2004},
volume = {22},
number = {2},
pages = {393},
author = {Tsagas, C. G.},
title = {Electromagnetic fields in curved spacetimes},
journal = {Classical and Quantum Gravity},
doi = {10.1088/0264-9381/22/2/011}
}
\bibliographystyle{unsrt}
\end{document}